# Demonstration of a three-dimensional current mapping technique around a superconductor in a prototype of a conventional superconducting fault current limiter

Md. Arif Ali, S. S. Banerjee*

Department of Physics, Indian Institute of Technology Kanpur, Kanpur - 208016, Uttar Pradesh, India

**Abstract**: Here we describe a three-dimensional current mapping technology developed for a superconductor using an array of Hall sensors distributed around it. We demonstrate this in a prototype similar to a conventional resistive superconducting fault current limiter (SCFCL). By calibrating the Hall sensor voltage, we can directly measure the distribution of the currents in the superconductor and the shunt. Using pulsed measurements, we measure the fractions of current distributed between the superconductor and shunt resistor parallel combination when a fault-like condition is mimicked in the system. Using the Hall array measurements, we generate a real-time three-dimensional map of local average current distribution around the superconductor used in our prototype of SCFCL. Our measurements show that, even for currents less than the critical current a non-uniform current flow pattern exists around the superconductor which we have used in the prototype. The capability of real-time, three-dimensional monitoring of the average local current distribution offers a way for the early detection of instabilities like hotspots developing in a superconductor. We discuss the use of this technique to not only show how it offers early detection and protection against instabilities developing in the superconductor, but also how it offers an added flexibility, namely, a user-settable fault current threshold.

*email: satyajit@iitk.ac.in

# SECTION I.

Introduction

The demands on modern power grids are complex as they need to handle both conventional and unconventional sources. These grids are prone to damage caused by fault currents due to one or more combinations of events like sudden large fluctuation in power on the demand or generation side, physical faults in the power distribution network due to extreme weather events, or other causes. The fault leads to damage of power grids, disrupts power distribution, and has a major economic fallout. To protect these power grids, fault current limiters are employed [1], [2], [3], [4], [5], [6], [7], [8]. Compared to conventional fault current limiter technologies, of particular interest in recent times has been the development of Superconducting Fault Current Limiters (SCFCL). The SCFCL [1], [2], [9], [10], [11], [12], [13], [14], [15], [16], provides an energy-efficient, automated and a cost effect solution to prevent damage to grids due sudden large fault current surges.

The impedance of the SCFCL is high when the fault-current ($I_{FL}$) exceeds a threshold value. The threshold current is set by the critical current $I_C$ of the superconductor used in SCFCL. The increased impedance restricts further increase of the current surge. However the same SCFCL under normal operating conditions, namely, $I_{FL} < I_C$, offers negligible impedance to the flow of current. An SCFCL operates as an automatic high current switch which swings between high and low resistance states depending on the value of the current. SCFCL's are of resistive [1], [17] and inductive type SCFCL [1], [18]. In this paper, our discussions will be related to the resistive type SCFCL. It is to be noted that in any SCFCL the $I_C$ is the critical parameter which determines the operating conditions of a SCFCL. The $I_C$ is completely determined primarily by the extent of vortex pinning strength in the superconducting material, which inturn depends on the material processing and synthesis procedures. The $I_C$ value also changes with magnetic field and operating temperature of the superconductor. In recent times, the use of high-temperature superconducting materials in SCFCL at temperatures below 115 K has allowed these SCFCL limiters to use cheap liquid Nitrogen as a cryogen for cooling the superconductor below its critical transition temperature, $T_C$. Such advances in SCFCL have happened due to the synthesis of high-temperature superconducting material in various forms like polycrystalline [19], thin-film [20], bulk [21], [22], wires[23], rings[24] and tapes [25] with high $I_C$. These high-temperature superconductors (HTSC) based SCFCL are made using thin-film, tapes, or wires of HTSC's materials like BSCCO-2212 ($Bi_2Sr_2CaCu_2O_{8+y}$), BSCCO-2223 ($Bi_2Sr_2Ca_2Cu_3O_{10+y}$) or YBCO ($YBa_2Cu_3O_{7+x}$). While SCFCLs are lucrative in their operation as fault limiting devices, there is a feature of these superconductors which leads to unpredictability in their operation.

One reason for the failure of the SCFCL during its operation is due to development of instabilities [26] like hotspots [27] generated in the superconductor used in the SCFCL. These hotspots are local high dissipation regions generated in the superconductor during the passage of a large current. The enhanced dissipation from these hotspot regions [28] results in a rapid increases in the size of the local high dissipating regions in the superconductor. This further enhances dissipation, which inturn rapidly causes local regions in the superconductor to turn normal. When this happens in the presence of high current flow, these local normal regions with high resistance produce more dissipation. This leads to a catastrophic increase in the amount of heat dissipated locally in the material, physically damages the superconducting material used in the SCFCL. Such a failure in the operation of SCFCL due to such instabilities happens even at operating currents, $I < I_{FL} < I_C$. While the source of the instabilities isn't fully understood, one possible reason for hotspot generation is local regions in the superconductor with lower $I_C$ than neighboring regions with higher $I_C$. These regions with locally suppressed $I_C$ turn normal, at $I$ less than the average $I_C$ of the superconductor. It is found that to reduce the probability of hotspot generation in the superconductors, the material needs to be placed in direct contact with the high conductivity metallic sheet (a metal shunt made from, say, Ag, Au). The metal shunt helps to smoothen out the irregularities of the temperature distribution along the length of the superconductor used in the SCFCL [29]. The limited reaction time of circuits to detect instabilities often prevents timely detection and intervention of a rapidly developing hotspot instability. Hence, there is a need to incorporate a sensor inside an SCFCL to allow real-time direct imaging of changes in current distribution occurring anywhere in the superconductor during the SCFCL's operation. Such a system offers a way for monitoring and warning of the development of instability in a superconductor. It is an essential component for ensuring the reliable and fail-proof operation of SCFCL.

We describe here the development of a conventional SCFCL with sensors that enable the real-time monitoring of the local current distribution at any location on the surface of the superconductor used in the SCFCL. We demonstrate this on a prototype that is similar to a conventional resistive SCFCL, with the addition of an array of low-temperature Hall sensors distributed around the entire superconductor [30]. The current flow generates a magnetic field distributed around the superconducting tube used in the SCFCL. The magnetic field distribution is measured by the array of Hall sensors and the information is used to generate a spatial map of the superconductor's current density distribution. Using this sensor array configuration, we are able to directly measure the amount of current distributing between the superconductor and the shunt before and after a fault like condition is produced. Our measurements at currents close to $I_C$ reveal the presence of non-uniform current distribution on the body of the superconductor used in our SCFCL prototype which evolve with current. The study suggests the presence of non-uniform vortex pinning in the superconductor. Our demonstration of continuous real-time monitoring and mapping of the average local current distribution across a macroscopic superconductor, offers a way for early detection of any instability developing in superconductors under high current situations. Such sensor arrays can also be deployed in the resistive SCFCL designs which are in operation in different application sectors. Here we also discuss a design based on the hall sensor array configuration implemented in SCFCL, which offers an added advantage. The advantage relates to the flexibility of the fault current threshold being set by a user in resistive SCFCL, instead of being a predetermined fixed parameter which is fixed by $I_C$ where $I_C$ is determined by the superconductor's material parameters. With this advantage one can set the fault current threshold limit to any value less than $I_C$, such that the fault limiting operation automatically commences well before the fault condition is reached, and the SCFCL reverts back to normal operation after the fault condition ceases.

**SECTION II.**

**Development of resistive superconducting fault current limiter**: We develop the prototype of a resistive superconducting fault current limiter using (a) a superconducting tube without any metal shunt in direct contact with the superconductor, (b) an array of cryogenic Hall sensors distributed around the superconductor, (c) a parallel metal shunt physically separated from the superconductor, (d) supporting copper (Cu) plate and (e) a cryostat in which the SCFCL has to be immersed, not shown in the image in Fig.1. We would like to mention we have used the above design of a prototype of a conventional resistive SCFCL to help in simplify the design of deploying Hall sensors around the superconductor. The distribution of Hall sensors we implement in our prototype can also be implemented in realistic SCFLC's in operation. The general principle of the technique we demonstrate in our prototype here will remain unchanged for hall sensor array deployed on any other modern day design of resistive SCFLC.

The superconductor (in Fig 1(a)) is a tube of BSCCO-2223 ($Bi_2Sr_2Ca_2Cu_3O_{10+x}$, from CAN Superconductors) with $T_C$ of 110 K. The $I_C$ of the superconductor at 77 K is 125 A. In our prototype, we have intentionally used a superconductor with low $I_C$ so that we can test the fault conditions inside the lab using currents generated from a high current power supply. The cylindrical BSCCO-2223 tube does not contain any cylindrical metallic base on which the superconducting material is placed. This is unlike a conventional SCFCL design, where the superconductor is usually placed on the base of a high conducting metallic sheet (we will discuss this issue later). The dimensions of the superconducting tube is given in TABLE I below.

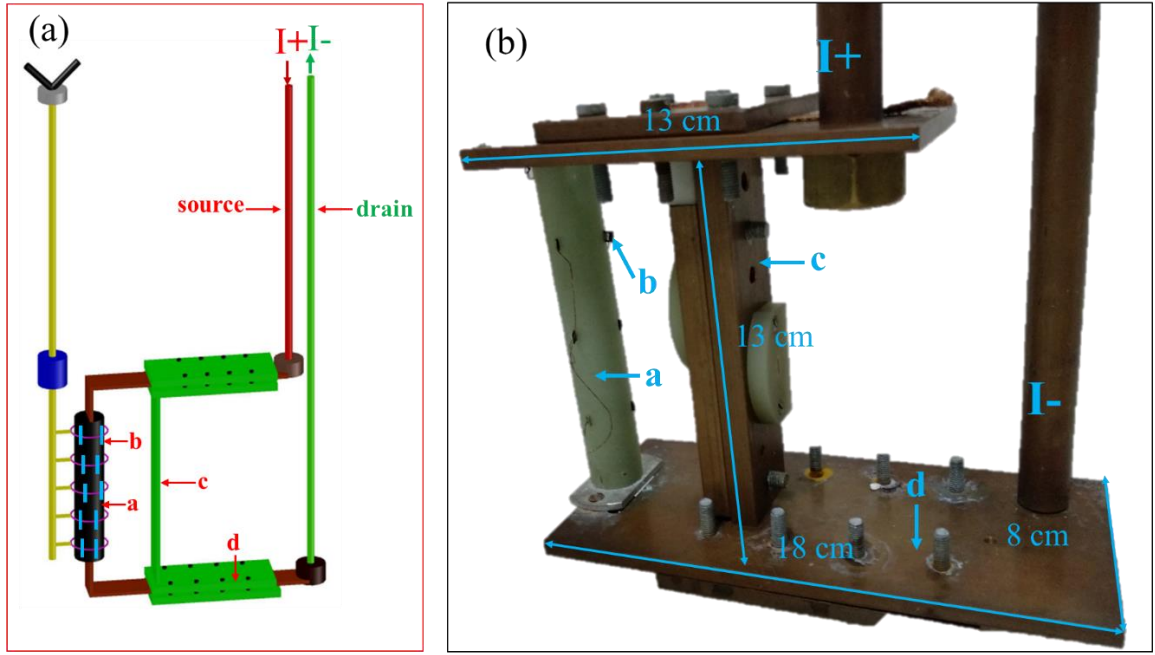

Fig. 1: (a) Schematic of the superconducting fault current limiter. The different components of the Superconducting Fault Current Limiter (SCFCL) where (a) superconductor, (b) Hall sensors around the superconductor, (c) copper shunt for bypassing the abnormal current, (d) supporting copper plate. (b) Image of the prototype of SCFCL with dimension as mentioned in the image.

The room temperature resistance of the superconductor as measured between the current-carrying copper leads connected to the superconductor is ~ 0.26 Ω. In our prototype, using the array of Hall sensors, we measure local magnetic field distributed across a large piece of superconductor. The length of the superconducting cylinder used in our design is 12 cm. To do these measurements across a wide surface area using the array of the Hall sensors, an essential requirement is to ensure that each Hall sensor in the array is positioned at the same height above the surface of the superconductor.

TABLE I

Parameters of Superconducting Tube

| Quantity | BSCCO-2223 |
|---|---|
| Shape | Bulk |
| Outer diameter of the cylinder (mm) | 10 |
| Inner diameter of the cylinder (mm) | 8 |
| Thickness of the cylinder (mm) | 1 |
| Length of the cylinder (mm) | 120 |
| Length of the copper leads/braids (mm) | 200 |

To ensure this, the Low-temperature cryogenic Hall sensors (see location marked b in Fig 1(a)) are placed around the superconductor by fixing them into the slits carved on the body of a G10 cylinder which is concentrically placed around the superconducting cylinder. The diameter of the G-10 cylinder allows gap of 3 mm between the outer surface of the superconducting cylinder and the inside surface of the G10 cylinder. This 3 mm gap ensures that sufficient cryogenic liquid is always present around the superconductor and prevents thermal instabilities from developing in the surface superconductor when high currents are flowing in the superconductor. We find that for negligible gap between the inner wall of the G-10 and outer wall of the superconducting cylinder, significant thermal instabilities arise in the superconductor during high current operation, which can also lead to the superconductor getting damaged. With these constraints and considerations, we found that the Hall sensors can be arranged at an optimal distance of 5 mm uniformly above the surface of the superconductor cylinder. We use a hollow superconducting cylinder, as it also allows the inside and the outer portions of the cylinder to be in contact with liquid Nitrogen. This helps in providing a larger surface area for cooling the superconductor. We find all of these are effective in minimizing any accidentally heating of the superconductor due to creation of thermal instabilities due to insufficient contact with cryogens and the superconductor, especially during the high current operation of the SCFCL prototype. The slits made in the G10 tube are ensure all the Hall sensors at a uniform distance of ~ $5 \pm 0.2$ mm from the surface of the superconductor. Here we would like to mention that groups in the past have measured the local magnetic field distribution of relatively small samples using the Hall probe technique [31, 32]. However the typical dimensions of the samples on which the local magnetic fields measured, were in the range of few millimeters. Here we use an array of Hall sensors to measure the local magnetic field distribution around a relatively large sized superconductor. The aim to show how such an array of Hall sensors can be also deployed in a modern resistive SCFCL design to measure the field distribution across a macroscopic sized superconducting material.

When current is sent into the cylindrical, hollow superconducting tube, the magnetic field around the tube has azimuthal symmetry, and it circulates concentrically around the cylinder. The strength of the field decays radially outwards from the axis of the cylinder. The Hall sensors are positioned in the G10 slits in such a way that the magnetic field around the tube hits the active area of the Hall sensor normally. This will be described in greater detail in the subsequent section on calibration. We use AlGaAs Hall sensors which are capable of working at liquid Helium temperature of 4.2 K, although in the present case, we operate down to liquid Nitrogen temperature of 77 K. The Hall voltage ($V_H$) from the Hall sensors is proportional to the magnetic field experienced by the sensors. The Hall sensors array is connected in series, and the $V_H$ is readout sequentially from each sensor in the array using a Keysight multichannel scanner cum digital Multifunctions Switch/Measure Unit (Model: 34980A). Our present design has seven Hall sensors distributed around the superconducting cylinder with the current inputs of all the Hall sensors connected in series. Typical parameters of the Hall sensors are given in supplementary information. For our Hall sensors, we use a current of 10 mA while measuring $V_H$. From the measured $V_H$ and the Hall coefficient ($R_H$) of the sensor, the magnetic field ($B$) experienced by the sensor is determined as $V_H = KB$. The typical value of $K$= 90 mV/T for one of the sensors. The active area of each Hall sensor ~ 100 μm × 100 μm with overall dimensions of 4 mm × 5 mm × 1 mm. Supporting copper plates (see location d in Fig. 1(a)) are used for making a sandwich-type electrical contact with the braided copper leads on the superconducting cylinder. The current is fed to the system using solid Cu rods (marked $I+$ and $I-$ in Fig.1) using high current carrying aluminum cables. The cryostat we have used is a home built with a doubled-walled chamber cryostat made from stainless steel. The outer chamber is pumped down to $10^{-4}$ mbar pressure using a turbo molecular pump. The cryostat holds up to 80 liters of liquid Nitrogen with a loss rate of 3 liters per hour. The entire SCFCL setup of Fig.1 is immersed in the cryostat.

Another aspect of our design includes a parallel metal shunt (see location c marked in Fig. 1(a)), which is a copper plate of dimension 13 cm × 3.2 cm × 1 cm. The shunt is used to bypass the current entering the superconductor in the event of any instability generated in the superconductor. In conventional SCFCL designs, a high electrically conducting metallic shunt is usually put in direct contact [33] with the superconducting material to ensure uniform heating of the superconductor and minimize the creation of hotspots. Instead of this configuration a normal conducting coil in parallel to the superconductor have also been used [34] to suppress hotspots. To further mitigate the hotspots problem, an additional shunt resistor with sufficiently low resistance, made of non-inductively wound wire, is connected in parallel with the YBCO thin film. In our prototype, we keep the shunt which is a rectangle block of copper physically separated from the SC cylinder (see Fig. 1(a) and (c)). Using a configuration where the shunt resistor and the superconductor are physically separate, have been used in some designs of resistive SCFCL's, while there also exists designs where the two are in direct physical contact. We believe the design where they are physically separate is advantageous, as it reducing the probability for generating thermal instabilities in the superconductor used in SCFCL (we will discuss this issue subsequently).

**SECTION III.**

**Calibration and conversion from field to current density**: Based on the Bio-Savart Law, for conductor carrying a uniform current density $J$, the magnitude of the azimuthal magnetic field, $B(r)$, at location $r$ (as shown in Fig. 2(a)) is directly proportional to $J$. One, can express a general relationship between the magnitudes of $B(r)$ and $J$ as, $B(r) = f(r).J$, where the shape of the finite-sized current-carrying conductor determines the explicit mathematical form of $f(r)$ and it also depends on the distance ($r$) from the surface of the conductor where the magnetic field $B$ is measured. With our calibration procedure described below, we determine $f(r)$ without explicitly needing its mathematical form. By using a predetermined value of $J$ sent through a conductor and by measuring $B(r)$ using Hall sensors placed at the location ($r$), the magnitude of $f(r)$ is determined using the above equation (see the location of multiple Hall sensor locations in Fig. 2(a) where $f(r)$ is determined at those locations). For our calibration, we have used a solid cylindrical Aluminium cable. We have also used a thin-walled cylindrical hollow copper tube which has similar dimensions as the superconducting cylinder. For a constant predetermined $J$ sent through these conductors, we determine the local field $B(r)$ at a fixed distance $r = 5.0 \pm 0.2$ mm from the surface of these conductors by placing Hall sensors inside slits of a G10 tube placed concentric to these conductors. The $J$ in these conductors is uniform. The entire setup is dipped in liquid Nitrogen to ensure the superconductor has always held a temperature of 77 K, which is below its superconducting transition temperature for all our experiments. After the above calibration procedure, we use these $f(r)$ values we have determined and the $B(r)$ measured around the superconductor to determine the $J$ using the above equation. We use this method to obtain maps of the spatial distribution of current density on the superconductor, which we will show in subsequent figures. We show in Fig. 2(b) measurement with a current ($I$) =225 A sent through a multicore Al cable (which is rated for a maximum 1000 A), and we measure $V_H$ from the sensors. Using the Hall coefficient of these sensors, we determine the magnetic field $B$ at the location of the Hall sensors above the Al cable. Figure 2(b) shows a plot of $B$ versus the current density ($J_{Al} = I/A$) where $A$ is the cross-sectional area = 110.85 mm$^2$ and $I$ is the current passing through the Al cable.

The $B$ measured by the Hall sensor would have the maximum contribution from the surface region of the current-carrying conductor as it is closest to the Hall sensor. The current density in this region is denoted as $J_{Al}$. The $B(J_{Al})$ curve we have determined above is nearly the same for the Hall sensors, which are placed around the superconducting cylindrical tube at a height of 5 mm from the surface of the superconductor. From Fig. 2(b) the value of $f(r) = m = 26$ G-mm$^2$/A. By sending in current through the superconducting tube and using the Hall sensors mounted in the G10 tube, the field $B_{SC}$ is measured at different locations on the superconductor. Using the $f(r)$, the measured $B_{SC}$ value is converted into an average local current density ($J_{SC}$) value which is present in a region of the superconductor just below the Hall sensor. As we can detect changes in local magnetic field with a resolution of ~ 0.25G, with a calibration factor $m = 26$ G-mm$^2$/A, the current density resolution obtained in our setup is 0.01 A.mm$^{-2}$. Along with Hall sensors placed around the superconductor, we also place Hall sensors 5 mm above the Cu shunt surface in order to measure the current flowing through the Cu shunt resistor. The above design can also be implemented in a high voltage situation. Under high voltage conditions, one may find an extra offset in the Hall voltage $V_H$ due to the electric field associated with the high voltages felt by the sensors.

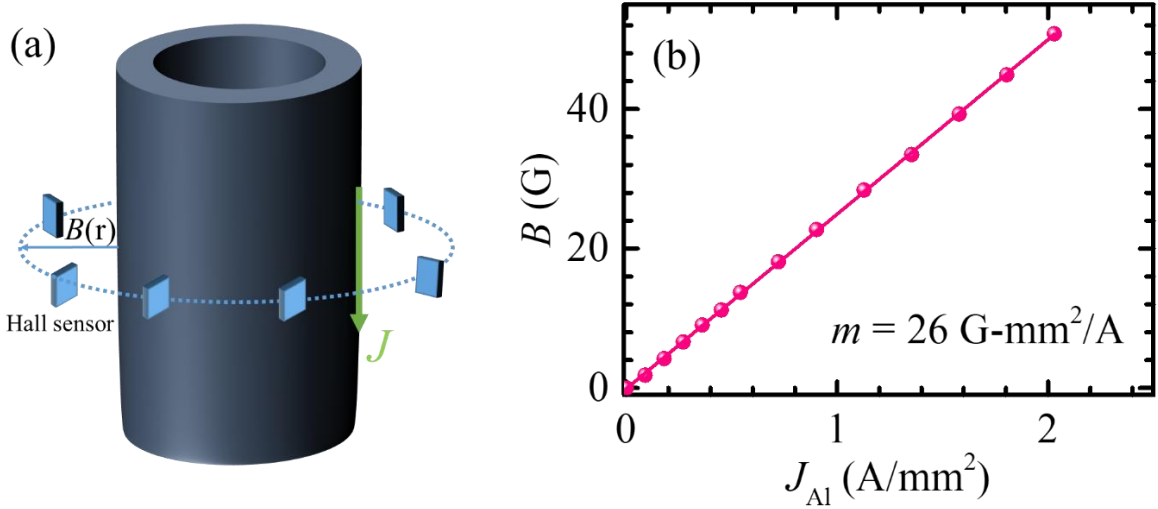

Fig. 2: (a) Schematic diagram shows Hall sensors have been placed around the superconductor. $B(r)$ is the magnitude of the azimuthal magnetic field at location $r$ is directly proportional to uniform current density $J$. (b) Magnetic field around the Aluminum cable ($B$) as a function of current density in cable ($J_{Al}$). The slope of the straight line ($m$) is 26 G-mm$^2$/A.

One can estimate this offset $V_H$ signal ($V_{H,off}$), by subtracting the $V_H$ measured around the superconductor carrying a current $I'$ under negligible applied voltage conditions and for the same $I'$ under the high applied Voltage ($V'$) applied in the vicinity of the Hall sensor, i.e., $V_{H,off}(I', V')=V_H((I', V') - V_H((I', \sim 0'))$. With this procedure, one will obtain $V_{H,off}$ for different applied voltage conditions. These offset voltages can be subtracted from the actual measured voltages from the sensors when operating the prototype under a high voltage environment.

The measurement of the voltage of the superconductor ($V$) versus $I$ of the superconductor used in our design is shown in Fig. 3(b). We determine the upper estimate of $I_C$ = 160 A for the superconductor, by fitting the $I$-$V$ curve with a well known form $V=V_0(I/I_C)^n$ [35] where $V_0 = 1$ µV/cm × $L$, is the mean baseline voltage corresponding to the electric field (E) criteria to determine $I_C$, namely, $E = 1$ µV/cm and $L$ is the distance between the voltage contacts ~ 10 cm, for our large superconductor. At $I = I_C$, $V = V_0$ and $V > V_0$ when $I > I_C$. The fit to this form is shown in the Fig.3(b) as the solid red curve through the data points. The fit gives a value of $n = 11$ and $I_C = 159 \pm 0.5$ A. The value of $n = 11 \pm 0.4$ is close to reported values of $n = 10$ obtained with similar fits to $I$-$V$ in BSCCO [36]. Apart from this procedure to determine $I_C$ we employ another criterion for determining $I_C$, namely, for $I \geq I_C$ the $V$ consistently deviates from $V_0$. Figure 3(b) (see arrow location) shows that with this criterion we get a lower $I_C = 125$ A. From the $I$-$V$ measurements we see that for the BSCCO sample used in our prototype, there is a spread of $I_C$ values between 125 A and 160 A, wherein the dissipation increases above the minimum threshold value, however beyond 160 A the dissipation increases rapidly. Therefore, superconducting sample used for the superconducting tube in our SCFCL prototype has an distribution of local $I_C$, where the average spread in the distribution is from 120 A to 160 A. The value of 160 A is the upper limit of $I_C$ distribution present in the sample and the mean $I_C$ is 140 A. The $I$-$V$ measurement was performed in our setup without connecting any shunt in parallel to the superconductor. Next, we introduce the copper shunt resistor in parallel to the superconductor as shown in the circuit diagram of Fig. 3(a). Here $I_{Sh}$ refers to the current through the shunt, $I_{SC}$ is the current through the superconductor and the total current $I_T = I_{SC} + I_{Sh}$. We measure the $I_{SC}$ and $I_{Sh}$ using the Hall sensors (as described earlier). Figure 3(c) shows that for $I_T < 150$ A, when $R$ of the SC is negligible, only about 24% of the current is flowing through the shunt and 76% ($I_{SC}$ ~ 114 A) is flowing through the SC. We call this regime I of operation of our SCFCL. Our direct measurement of current division shows that although the resistance of the superconductor is negligible (since $I_{SC}$ is less than the average $I_C$ of the superconductor), there is still about 20 % of the total current flowing through the shunt. This is in contrast to the conventional expectation of zero current to flow through the shunt when $I < I_C$. This observed current division between the shunt and the superconductor for $I < I_C$ can only be explained by the presence of finite differences in the contact resistance along the two parallel current paths, namely, one current path through the superconductor and another current path through the shunt, respectively. In this low current regime, the contact resistance differences along the two parallel current paths lead to the observed current distribution between the shunt and the superconductor although the $I_{SC} < I_C$.

In the high current regime greater than 100 A, all our measurements are done in a pulsed mode where $I$ is switched on, Hall sensor voltages are measured and the $I$ is then switched off. The current is kept on for few 10's of ms (~ 50 ms). This helps prevent heating-related damage to the superconductor. Above 100 A, the data are shown in Fig. 3(c) correspond to the transient measurement of $V_H$ from the sensors around the superconductor and shunt in order to detect possible switching of current paths. Above $I_T$ ~ 150 A, our transient measurements using the Hall sensors show (see Fig. 3(c)) that more current begins to flow through the shunt compared to the superconductor. We call this regime II of operation of our SCFCL. The behavior we see in regime II is different from that seen below 150 A in regime I. From Fig. 3(c) it appears that above $I_T = 150$ A significant amount of the current diverts out from the superconductor and into the shunt. Note that at $I_T = 150$ A the $I_{SC}$ is around 114 A.

The only way this behavior can happen is, if the resistance of the superconductor, including the contact resistance, exceeds the resistance of the shunt (inclusive of the contact resistance). This implies that an $I_{SC} = 114$ A exceeds the local $I_C$ in the superconductor. This is consistent with our estimate from Fig. 3(b) that the superconductor has a wide spread of $I_C$'s. While Fig. 3(b) shows that that the measured average spread of the $I_C$ in the sample is between 120 A to 160 A. This measured width corresponds the average width of the distribution of $I_C$ in the superconductor. Tails of the distribution of $I_C$ in the superconducting sample may extend to slightly below 120 A and slightly above 160 A, and these values which may not be detectable in an $I$-$V$ type of measurement as their fraction . Thus there will be local regions (albeit small fraction of the sample) which possess a $I_C$ slightly lower than 120 A. The $I_T$ at which $I_{SC}$ in the superconductor exceeds the local $I_C$ results in the superconductor resistance to rise rapidly here, resulting in the current getting diverted from the SC into the shunt. At this point of the momentary quench undergone by the superconductor, the $R$ of the superconductor at 77 K was measured to be ~ 0.17 $\Omega$. With this increase in resistance on the superconducting arm, the current will divert from the superconductor to the shunt resistor. From Fig. 3(c), we see from our measurements above 100 A we see that at $I_T = 200$ A only ~ 20% of $I_T$ current flows through the SC (~ 40 A < $I_C$) while 80% of $I_T$ flows through the shunt (~ 160 A). The momentary diversion of current results in a decrease of current flowing through the superconductor to below the $I_C$ value, with most of the current flowing through the shunt. In an actual SCFCL, this diversion of current into the shunt would help in limiting the fault current. In our prototype, we are able to measure exactly the current distribution between the superconductor and shunt resistor in the prototype of an SCFCL, especially close to situations similar to a fault condition. We mimic the fault condition in our experiment by sending in currents larger than $I_C$ of the superconductor.

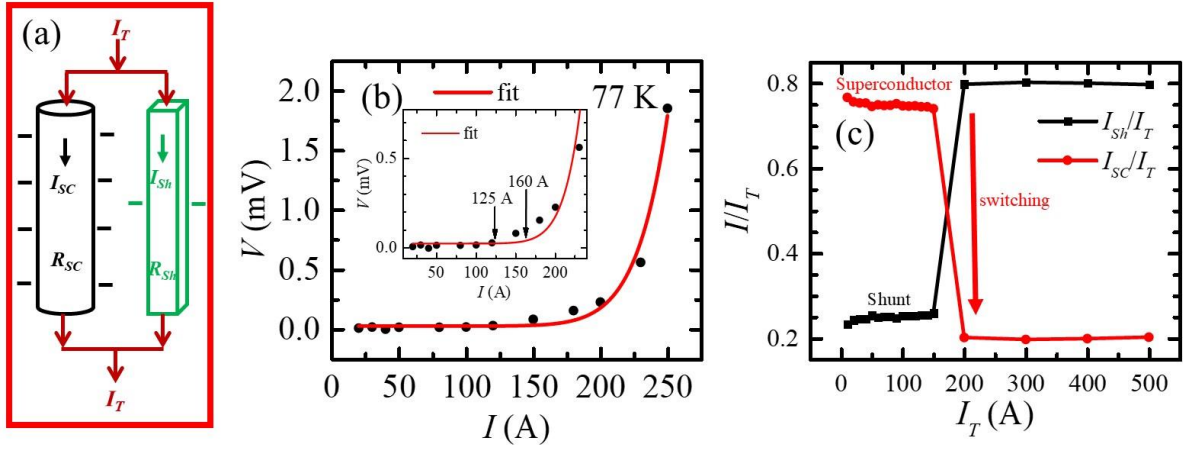

Fig. 3: (a) Schematic of the parallel configuration of shunt (Sh) and superconductor (SC) used while for the fault current operation in Fig.3(c). Around the superconductor and the shunt, the horizontal dashed lines represent Hall sensors used to measure the current distribution. (b) Measurement of voltage of the superconductor ($V$) versus current ($I$) through it. For this measurement, no shunt was connected in parallel with the SC. (c) Using the parallel combination of Sh and SC we measure using the Hall sensors the current flowing through the Sh and SC. Using this, we determine the ratio $I_{Sh}/I_T$ and $I_{SC}/I_T$. (see text for details).

The contact resistance along the $I_{SC}$ path leads to a diversion of current into the shunt even in the normal working condition. Typically, in any SCFCL, during a non-fault operation conditions, the shunt resistor value should be greater than the contact resistance between the current-carrying copper leads with the superconductor. This condition ensures that maximum current passes through the superconductor in the non-fault condition, and no current flows through the shunt. When a fault occurs as the resistance on the superconducting path increases beyond the resistance on the shunt path, hence the current is diverted through the shunt from the superconductor. This situation also necessitates that the resistance on the shunt path should be low resistance, so that it can to provide a low dissipation shunt path for the high fault current.

Although a superconductor has zero resistance for currents less than its critical current, there is always an ever present finite contact resistance at the joint between the current-carrying copper leads and the superconductor. This contact resistance is ever present. In our setup at 77 K the contact resistance of the superconductor and the current-carrying copper leads is ~ 170 m$\Omega$. This net resistance, along with the shunt resistor path (= shunt resistance plus the contact resistance of the current leads with the shunt resistor), should be well above ~ 170 m$\Omega$, for ensuring that all the current passes through the superconductor during normal operation of our SCFCL prototype (i.e. when $I < I_C$). However, the resistance along the shunt path cannot be increased significantly as the dissipation also increases as $I^2 R_{Sh}$, where $R_{Sh}$ is the net resistance along the shunt path. Even a low dissipation shunt resistance $R_{Sh} = 1$ $\Omega$ at DC current of 100 A, leads to a large dissipation of 10 kW. To enable low dissipation along with low contact resistances plus low shunt resistance, we achieved a value of 500 m$\Omega$ along the shunt resistance path. As the resistance along the shunt and superconducting paths are comparable, this leads to the observation of a current division between the superconductor and shunt at low $T$, even for $I < I_C$ operation of our SCFCL prototype. Here we would like to comment that in a number of resistive SCFCL's the superconducting material is coated with metallic shunt layer. In this situation, it is not possible to figure out if any current is diverting through the shunt due to any imbalance in the contact resistance along the superconducting path and the shunt path (as discussed above).

Based on this observation, we believe that similar current division would also occur in conventional SCFCL system designs wherein a superconductor is usually kept in direct physical contact with a low resistance metal shunt (the superconductor is placed directly on top of the shunt) in order to ensure uniform heating and control hotspots. In such designs, even in normal operation, a significant amount of current could be flowing through the metallic shunt due to imbalances in contact resistance (as discussed for our design). The current through the shunt would act as a source of Joule heating which in turn would generate thermal instability in the superconductor, which is in proximity with the shunt in these designs. It should be mentioned that in the designs where the superconductor and shunt are placed in direct physical contact, it is almost impossible to measure the extent of current distribution between the superconductor and shunt as hall sensors cannot be placed separately on the shunt and the superconductor to measure the currents in the shunt and superconductor. For designs where the shunt and superconductor are physically separate, one is able to measure the division of current flow between the shunt and superconductor. Hence conventional SCFCL system where the superconductor is kept in direct contact with a metallic shunt has limited efficiency for preventing hotspots. Due to the above consideration, we believe that designs where the metallic shunt physically separate from the superconductor are less likely to generate hotspots due to dissipation in the shunt.

**SECTION IV.**

**Current density mapping**: In our superconducting fault current limiter, an array of Hall sensors have been placed around the superconductor as shown in Fig. 4(a) and described earlier. Figure 4(b) shows the ($z$, $\Phi$) coordinate for the location of the Hall sensors. The $z$-is the length measured from the bottom of the cylinder along its vertical axis and $\Phi$ is the azimuthal angle. By rotating the G10 cylinder, we obtain measurements at different $\Phi$. The rotation is by 10° of the G10 cylinder. We are unable to do a full 360° rotation for each sensor, as the wirings of the hall sensors get strained with rotation and often the currents and voltage leads on the sensors can break. The spatial resolution in our map is 5 ± 1 mm.

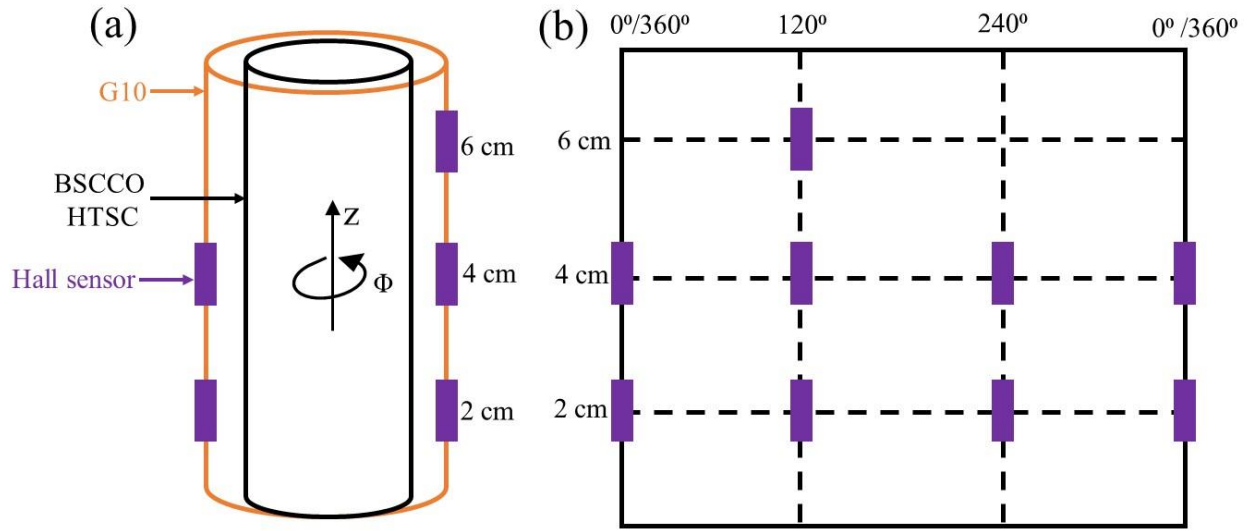

Fig. 4: (a) An array of seven Hall sensors have been kept around the superconductor. (b) Coordinates of the Hall sensors on the curved surface of the superconducting cylinder. Due to periodic boundaries, we repeat the sensors located along the 0°/360° line. The magnetic field intensity with the help of these coordinates of Hall sensors is used to monitor the system's instability. The length of the superconductor tube is 12 cm. Copper covers the top 2 cm of the superconductor leads joining the superconductor. The copper also covers 2 cm of the lower portion of the superconductor leads joint. The remaining 8 cm between the leads is the surface of the superconductor. The z = 2 cm measurement is from where the copper leads ending on the superconductor (which is about 4 cm from the base of the superconductor). Thus the hall sensors span across more than half the length of the superconductor cylinder.

Using this Hall sensor array configuration, we perform a real-time mapping of the current density in the superconducting tube. We perform these measurements by sending in 100 A of current in the superconductor. For this measurement, we do not connect any parallel Cu shunt to ensure 100% current flow into the superconductor. The $V_H$ from each Hall sensor in the array is periodically measured using a scanner - digital multimeter, which is interfaced to a computer. Here the measured $V_H$ from each sensor is converted into the $B$ experienced by the sensor at that location. The computer screen displays as a function of time the $B$ experienced by each of the Hall sensors distributed around the superconductor. This monitoring allows real-time detection of changes in $B$ due to instabilities generated in the superconductor. It is more convenient to look at all the fields measured by the Hall sensors as a map of $B$ as a function of $z$ and $\Phi$.

The measured $B$ from each Hall sensor in the seven sensors array is used to construct the $B$ $(z, \Phi)$ map distributed around the current-carrying superconductor (see Fig. 5(a)-(b)). The colour scale is from 23.84 G to 27.82 G of Fig. 5(a) when 100 A of current is sent through the superconductor. The deepest blue shade represents 23.84 G while the deepest red corresponds to 27.82 G. The portion of the plot with no colour (white) is the $(z, \Phi)$ regions that do not have any sensors. The plot shows the field distribution is not uniform around the superconductor, and we are having this for two different external current 100 A and 120 A. Using the $B(J_{Al})$ plot discussed in Fig. 2(b), the $B$ $(z, \Phi)$ map is converted into a three-dimensional $J_{SC}$ spatial map in the superconductor. Figure 5(c) shows the $J_{SC}(z, \Phi)$ map. The dark solid lines in Fig. 5(c) represent regions of constant $J_{SC}$ along the superconductor.

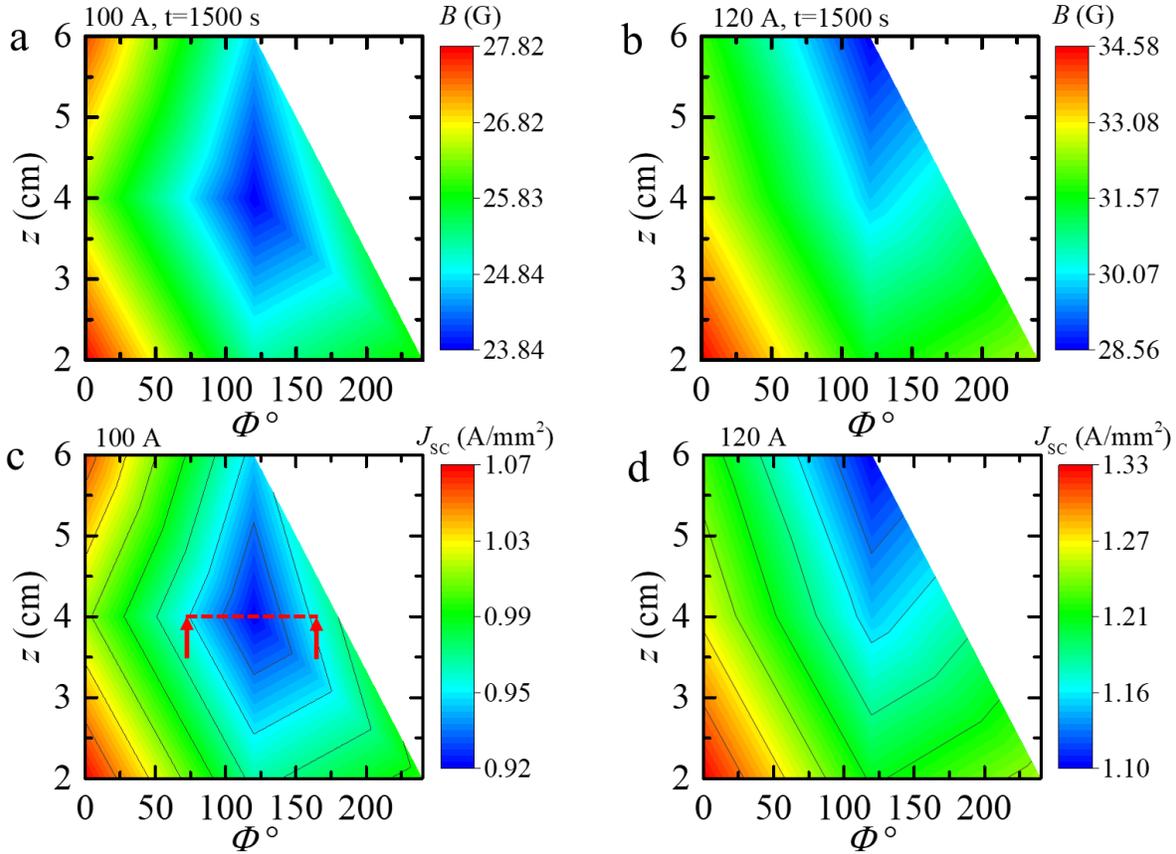

Fig. 5: (a)-(b) Real time magnetic field monitoring with seven Hall sensors around the superconductor for applied current 100 A and 120 A. (c)-(d) corresponding current density monitoring with seven Hall sensors around the superconductor. All the above measurements have been performed at 77 K, after 1500 seconds after sending in current into the superconductor.

Along this red dashed line marked with arrows in Fig. 5(c) we see that between $\Phi = 75°$ and $175°$ there is a region carrying significantly low $J_{SC}$ compared to regions below it. The map shows that although $I = 100$ A is lower than the $I_C$ (~ 120 A) of the superconductor, the $J_{SC}$ around the superconductor is intrinsically non-uniform. We have measured this current density distribution after 1500 s of passing 100 A current through the superconductor, and hence it represents a near steady-state current distribution in the superconductor. The local current density map clearly shows the presence of high current density and low current density regions in the SC. While we do not fully understand the reason for the non-uniform current distribution around the superconducting cylinder, we believe there is non-uniformity of pinning across the superconductor. Infact in from the *I-V* measurement in Fig.3(b), we see evidence for a distribution of $I_C$ in the sample, where the average width of the distribution ranges approximately from 120 A to 160 A. We believe there are regions where the local critical current is much lower than the bulk of the superconductor, especially in the region between $\Phi = 75°$ and $175°$ in Fig. 5(c). Due to this, the weakly pinned vortices begin moving in these regions in the presence of current. Hence in these weak pinning locations, the average local resistance of the superconductor is higher (due to the dissipating moving vortices) compared to neighbouring regions where vortices are strongly pinned and localized. Hence currents avoid these regions leading to a lower local *J* and pass through neighbouring regions. In Fig. 5(b), we show a magnetic field map when $I = 120$ A is sent through the superconductor. Figure 5 (d) shows the corresponding current density map. We see that at higher current $I = 120$ A in Fig. 5(d) that the region with suppressed *I* (bluish region) moves out from the field of view of the region mapped by the array of hall sensors. The region of suppressed current density is a region of instability (presumably a hot spot region) in the superconductor. This region dynamically evolves with higher current drives as seen in Fig.5(d). Note that in our superconducting sample which exhibits an average spread of $I_C$ between 120 A to 160 A, the mean $I_C \sim 140$ A. Thus our observation shows the presence of instability regions which dynamically evolve in the superconductor even at *I*, which are less than or close to the average critical current of the superconductor. Thus in our system offers real-time monitoring of $J(z, \Phi)$ around the superconductor. Detection of changes in the $J(z, \Phi)$ maps due to hotspot generation becomes convenient with our setup. The detection of hotspots also allows preventive action to avoid further evolution of this instability in the superconductor. Using a Cu shunt in parallel with the superconductor will enable the diversion of current whenever hotspots develop in the superconductor. The diversion of current into the shunt can be implemented by using a thyristor-based switch. This proposal we describe below.

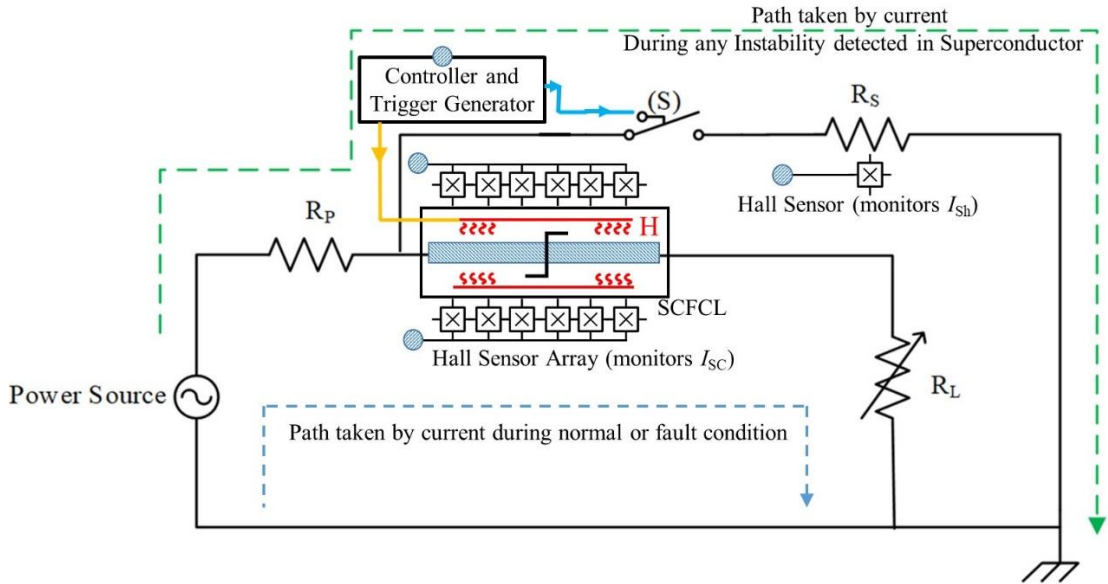

Fig. 6: The circuit consists of a power source, a protecting resistance $R_P$ for the source, a superconductor (blue hashed region), and load resistance $R_L$. H are heaters around the superconductor. The Hall sensors are shown around the superconductor. S is a switch, and $R_S$ is the shunt resistor. The control and trigger generator box shown in the schematic is an electron circuit that sends or switches off the current to the heater H and also controls

the opening and closing of S. The control and trigger circuit's response is determined by the inputs from the Hall sensor signals distributed around the superconductor.

## SECTION V.

**A proposal of an SCFCL design with user-settable fault current threshold**: It is common knowledge that if a portion of the superconductor is heated to above its superconducting transition temperature $T_C$ its resistance will increase. We make use of this simple principle in our proposal. In Fig. 6, we propose an SCFCL design [30] in which one can set the fault current limit ($I_{FL}$) at any current value which is less than the bulk $I_C$ of the superconductor, viz., $I_{FL} \leq I_C$. This is a useful option to have if one wants the SCFCL to begin its fault limiting operation right at the initial stages of the fault condition is setting in, i.e., when the fault current has increased beyond a certain value, yet it is still less than $I_C$ of the superconductor. In Fig. 6, we show the SCFCL, which consists of a superconductor with the Hall sensor array around it (as discussed above). The superconductor is connected in parallel to the shunt resistor ($R_S$) via a switch (S), and the superconductor is in series with the load resistance ($R_L$). Additionally, there are resistive heaters H could be coils of wire wound around some portions of the superconductor, which will heat these portions of the superconductor to above $T_C$. Typically, H could be resistance heater wire like Nichrome wire. The current flow in the heater H is controlled via a 'control and trigger generator circuit' shown inside a box in Fig. 6, which receives inputs from the Hall sensors, is used to control the opening and closing of switch S as well as the turning on or off the current flowing through H. During normal operation (non-fault condition), currents flow through the superconductor and the load $R_L$. During this normal operation, the switch (S) is maintained in an open condition. Also, no current is sent through the heater *H*. Once the measured Hall sensors signal crosses a preset value $I_{FL}$ (due to a fault condition), the controller and trigger circuit senses this and sends current into the heater wires H and simultaneously also turns the switch S into closed condition. As H heats up the portions of the superconductor around which it is wound to above $T_C$ the superconductor enters the high resistance state. With this, the current is immediately diverted into the parallel shunt ($R_S$) from the superconductor. Thus at any user-settable $I_{FL}$ value $< I_C$, one can use the above to cause switching of the current path from the superconductor into the shunt to begin a fault limiting operation. Once the fault condition ceases, the control and trigger circuit switches off the current in H and opens S. Usually, $I_C$ is predetermined by the material properties of the superconductor used in the SCFCL. Namely, it depends on the presence of effective strong pinning centers in the superconductor. Since $I_C$ is fixed by the material employed, hence in SCFCL, the fault operation happens only at $I_{FL} = I_C$, and the user has no freedom to set the $I_{FL}$. In the above implementations, the fault operation can be started at an $I_{FL}$ value that is less than $I_C$. We describe the above operation, with an example. In the design described above one can set for the SCFCL an $I_{FL} = xI_C$, where $x$ is less than or equal to 1. A value of $x = 1$ is the conventional threshold used in any SCFCL. In our proposed design, consider for example a user-settable fault threshold is set with $x = 0.7$, i.e., $I_{FL} = 0.7\ I_C$. The controller and trigger pulse generator in the system has been pre-programmed to generate a trigger pulse whenever the Hall sensor array detects a current value in excess of $0.7I_C$ flowing in the superconductor. The trigger pulse which is generated, activates two circuits, one circuit controls the switch S, and the other circuit controls the heaters around the superconductor. Thus, when a fault appears, and the current increases and crosses $0.7I_C$, the controller and trigger pulse generator produce a trigger pulse which closes the switch S. This results in the current diverting through the shunt resistor. Simultaneously the trigger pulse also activates the heater circuit and the heaters around the superconductor heat up the superconductor to drive it into a high resistance state. Thus for all fault current $I > 0.7I_C$, the current will be forcibly diverted along the $R_P$ and $R_S$ path (see green dashed path in Fig. 6), instead of the current flowing through $R_P$, superconductor, and the $R_L$ path (blue dashed path in Fig. 6). Once the fault condition passes away, and the Hall sensor around $R_S$ detects that $I$ flowing in the shunt has fallen well below $0.7\ I_C$, the controller and trigger pulse generator system will activate another pulse which will cause the switch S to open and also simultaneously deactivate the heater circuits. The superconductor will be cooled down to below $T_C$, and once again, the current will flow through the $R_P$, superconductor, $R_L$ path (see the blue dashed path in Fig. 6). While in conventional SCFCL, the fault limiting action begins at $I = I_C$, i.e., for $x = 1$, in our proposed design, the user has the flexibility to initiate the fault limiting operation at any $x \leq 1$. Infact here, the superconductor is prevented from being driven repeatedly into the high dissipation state at $I_C$ whenever a fault condition occurs. The switch $S$ also serves as additional protection for the SCFCL. Suppose at any $I$ the Hall sensors detect any unusual distribution of currents in the superconductor setting due to some instability. In that case, switch $S$ can be triggered into its closed position to protect the superconductor from any possible damage. Such a system will offer fail-safe, long-term operation of the SCFCL.

**Conclusion:** Our work shows that it is possible to incorporate an array of Hall sensors around the superconductor used in an SCFCL. The Hall sensors allow for real-time monitoring, and mapping of current density around the superconductor is possible. This real-time monitoring offers new possibilities of early detection of instability developing in the superconductor like hotspot generation. Our SCFCL system provides enhanced protection against damage and allows for a reliable operation. With our superconducting fault current limiter design, we get the flexibility of fault limiting operation, which can be set at any fault current limit threshold predetermined by the user. Such flexibility in operation has never been offered before in earlier designs of superconducting fault current limiters.

**SECTION VI.**

**Acknowledgements**: S. S. Banerjee would like to acknowledge IIT Kanpur and DST-AMT, the Government of India, for funding this project. Md. Arif Ali thanks IIT Kanpur for support.

**DATA AVAILABILITY**: The data that support the findings of this study are available from the corresponding author upon reasonable request.

# Demonstration of a three-dimensional current mapping technique around a superconductor in a prototype of a conventional superconducting fault current limiter

Md. Arif Ali, S. S. Banerjee*

Department of Physics, Indian Institute of Technology Kanpur, Kanpur - 208016, Uttar Pradesh, India

The following supplementary section gives the information about the Hall sensors we have used in our experiments.

TABLE I

Specifications of HHP-MU series Hall sensors from AREPCO S.R.O.

| Parameter | Unit | 297 K | 77 K |
|---|---|---|---|
| Nominal control current | mA | 10 | 10 |
| Maximum control current | mA | 12 | 15 |
| Sensitivity at $I_n$ | mV/T | 90.5 | |
| Linearity error up to 1 T | % | < 0.2 | |
| Change of sensitivity due to reversing of the magnetic field | % | < 1 | |
| Active area dimension | μm | $100 \times 100$ | |
| Overall dimension (w × l × h) | mm | $4 \times 5 \times 1$ | |

TABLE II

Specifications on sensitivity and offset voltage of HHP-MU series Hall sensors from AREPCO S.R.O.

| Product Number | Sensitivity at $I_n$ at 77 K (mV/T) | Offset voltage at $I_n$ at 77 K (μV) |
|---|---|---|
| 1749 | -50 | 97.3 |
| 1750 | 114 | 91.6 |
| 1751 | -56 | 91.1 |
| 1752 | 54 | 92.3 |
| 1753 | 107 | 93.5 |
| 1754 | 62 | 90.5 |
| 1755 | -85 | 79.8 |